\documentclass[longauth]{aa}
\usepackage{graphicx}
\usepackage{txfonts}
\usepackage{xcolor}
\usepackage{hyperref}
\hypersetup{colorlinks, linkcolor=blue, citecolor=blue, urlcolor=blue}

\newcommand{\te}{t_{\rm E}}
\newcommand{\thetae}{\theta_{\rm E}}

\newcommand{\pie}{\pi_{\rm E}}

\newcommand{\dl}{D_{\rm L}}
\newcommand{\ds}{D_{\rm S}}

\definecolor{brown}{rgb}{0.59, 0.29, 0.0}
\definecolor{darkgreen}{rgb}{0.0, 0.42, 0.24}
\definecolor{darkblue}{rgb}{0.01, 0.31, 0.59}
\definecolor{darkblue}{rgb}{0.0, 0.25, 0.42}
\definecolor{blue}{rgb}{0.0,0.0,1.0}
\definecolor{green}{rgb}{0.0,1.0,0.0}

\begin{document}

\title{
KMT-2021-BLG-0284, KMT-2022-BLG-2480, and KMT-2024-BLG-0412: Three
microlensing events involving two lens masses and two source stars
}
\titlerunning{KMT-2021-BLG-0284, KMT-2022-BLG-2480, KMT-2024-BLG-0412}

\author{
     Cheongho~Han\inst{\ref{inst1}} 
\and Andrzej~Udalski\inst{\ref{inst2}} 
\and Ian~A.~Bond\inst{\ref{inst3}}
\and Chung-Uk~Lee\inst{\ref{inst4},\ref{inst25}} 
\and Andrew~Gould\inst{\ref{inst5},\ref{inst6}}      
\\
(Leading authors)
\\
     Michael~D.~Albrow\inst{\ref{inst7}}   
\and Sun-Ju~Chung\inst{\ref{inst4}}      
\and Kyu-Ha~Hwang\inst{\ref{inst4}} 
\and Youn~Kil~Jung\inst{\ref{inst4}} 
\and Yoon-Hyun~Ryu\inst{\ref{inst4}} 
\and Yossi~Shvartzvald\inst{\ref{inst8}}   
\and In-Gu~Shin\inst{\ref{inst9}} 
\and Jennifer~C.~Yee\inst{\ref{inst9}}   
\and Hongjing~Yang\inst{\ref{inst10}}     
\and Weicheng~Zang\inst{\ref{inst9},\ref{inst10}}     
\and Sang-Mok~Cha\inst{\ref{inst4},\ref{inst11}} 
\and Doeon~Kim\inst{\ref{inst1}}
\and Dong-Jin~Kim\inst{\ref{inst4}} 
\and Seung-Lee~Kim\inst{\ref{inst4}} 
\and Dong-Joo~Lee\inst{\ref{inst4}} 
\and Yongseok~Lee\inst{\ref{inst4},\ref{inst11}} 
\and Byeong-Gon~Park\inst{\ref{inst4}} 
\and Richard~W.~Pogge\inst{\ref{inst6}}
\\
(The KMTNet Collaboration)
\\
     Przemek~Mr{\'o}z\inst{\ref{inst2}} 
\and Micha{\l}~K.~Szyma{\'n}ski\inst{\ref{inst2}}
\and Jan~Skowron\inst{\ref{inst2}}
\and Rados{\l}aw~Poleski\inst{\ref{inst2}} 
\and Igor~Soszy{\'n}ski\inst{\ref{inst2}}
\and Pawe{\l}~Pietrukowicz\inst{\ref{inst2}}
\and Szymon~Koz{\l}owski\inst{\ref{inst2}} 
\and Krzysztof~A.~Rybicki\inst{\ref{inst2},\ref{inst8}}
\and Patryk~Iwanek\inst{\ref{inst2}}
\and Krzysztof~Ulaczyk\inst{\ref{inst12}}
\and Marcin~Wrona\inst{\ref{inst2},\ref{inst24}}
\and Mariusz~Gromadzki\inst{\ref{inst2}}          
\and Mateusz~J.~Mr{\'o}z\inst{\ref{inst2}} 
\\
(The OGLE Collaboration)
\\
     Fumio~Abe\inst{\ref{inst13}}
\and Richard~Barry\inst{\ref{inst14}}
\and David~P.~Bennett\inst{\ref{inst14},\ref{inst15}}
\and Aparna~Bhattacharya\inst{\ref{inst13},\ref{inst14}}
\and Hirosame~Fujii\inst{\ref{inst13}}
\and Akihiko~Fukui\inst{\ref{inst16},}\inst{\ref{inst17}}
\and Ryusei~Hamada\inst{\ref{inst18}}
\and Yuki~Hirao\inst{\ref{inst18}}
\and Stela~Ishitani Silva\inst{\ref{inst15},\ref{inst19}}
\and Yoshitaka~Itow\inst{\ref{inst13}}
\and Rintaro~Kirikawa\inst{\ref{inst18}}
\and Naoki~Koshimoto\inst{\ref{inst20}}
\and Yutaka~Matsubara\inst{\ref{inst13}}
\and Shota~Miyazaki\inst{\ref{inst18}}
\and Yasushi~Muraki\inst{\ref{inst13}}
\and Greg~Olmschenk\inst{\ref{inst14}}
\and Cl{\'e}ment~Ranc\inst{\ref{inst21}}
\and Nicholas~J.~Rattenbury\inst{\ref{inst22}}
\and Yuki~Satoh\inst{\ref{inst18}}
\and Takahiro~Sumi\inst{\ref{inst18}}
\and Daisuke~Suzuki\inst{\ref{inst18}}
\and Mio~Tomoyoshi\inst{\ref{inst18}}
\and Paul~J.~Tristram\inst{\ref{inst23}}
\and Aikaterini~Vandorou\inst{\ref{inst14},\ref{inst15}}
\and Hibiki~Yama\inst{\ref{inst18}}
\and Kansuke~Yamashita\inst{\ref{inst18}}
\\
(The MOA Collaboration)
}

\institute{
      Department of Physics, Chungbuk National University, Cheongju 28644, Republic of Korea\label{inst1}                                                          
\and  Astronomical Observatory, University of Warsaw, Al.~Ujazdowskie 4, 00-478 Warszawa, Poland\label{inst2}                                                      
\and  Institute of Natural and Mathematical Science, Massey University, Auckland 0745, New Zealand\label{inst3}                                                    
\and  Korea Astronomy and Space Science Institute, Daejon 34055, Republic of Korea\label{inst4}                                                                    
\and  Max Planck Institute for Astronomy, K\"onigstuhl 17, D-69117 Heidelberg, Germany\label{inst5}                                                                
\and  Department of Astronomy, The Ohio State University, 140 W. 18th Ave., Columbus, OH 43210, USA\label{inst6}                                                   
\and  University of Canterbury, Department of Physics and Astronomy, Private Bag 4800, Christchurch 8020, New Zealand\label{inst7}                                 
\and  Department of Particle Physics and Astrophysics, Weizmann Institute of Science, Rehovot 76100, Israel\label{inst8}                                           
\and  Center for Astrophysics $|$ Harvard \& Smithsonian 60 Garden St., Cambridge, MA 02138, USA\label{inst9}                                                      
\and  Department of Astronomy, Tsinghua University, Beijing 100084, China\label{inst10}                                                                            
\and  School of Space Research, Kyung Hee University, Yongin, Kyeonggi 17104, Republic of Korea\label{inst11}                                                      
\and  Department of Physics, University of Warwick, Gibbet Hill Road, Coventry, CV4 7AL, UK\label{inst12}                                                          
\and  Institute for Space-Earth Environmental Research, Nagoya University, Nagoya 464-8601, Japan\label{inst13}                                                    
\and  Code 667, NASA Goddard Space Flight Center, Greenbelt, MD 20771, USA\label{inst14}                                                                           
\and  Department of Astronomy, University of Maryland, College Park, MD 20742, USA\label{inst15}                                                                   
\and  Department of Earth and Planetary Science, Graduate School of Science, The University of Tokyo, 7-3-1 Hongo, Bunkyo-ku, Tokyo 113-0033, Japan\label{inst16}  
\and  Instituto de Astrof{\'i}sica de Canarias, V{\'i}a L{\'a}ctea s/n, E-38205 La Laguna, Tenerife, Spain\label{inst17}                                           
\and  Department of Earth and Space Science, Graduate School of Science, Osaka University, Toyonaka, Osaka 560-0043, Japan\label{inst18}                           
\and  Department of Physics, The Catholic University of America, Washington, DC 20064, USA\label{inst19}                                                           
\and  Department of Astronomy, Graduate School of Science, The University of Tokyo, 7-3-1 Hongo, Bunkyo-ku, Tokyo 113-0033, Japan\label{inst20}                    
\and  Sorbonne Universit\'e, CNRS, UMR 7095, Institut d'Astrophysique de Paris, 98 bis bd Arago, 75014 Paris, France\label{inst21}                                 
\and  Department of Physics, University of Auckland, Private Bag 92019, Auckland, New Zealand\label{inst22}                                                        
\and  University of Canterbury Mt.~John Observatory, P.O. Box 56, Lake Tekapo 8770, New Zealand\label{inst23}                                                      
\and  Villanova University, Department of Astrophysics and Planetary Sciences, 800 Lancaster Ave., Villanova, PA 19085, USA\label{inst24}                          
\and  Corresponding author\label{inst25}                                                                                                                           
}                                                                                                                                                       
\date{Received ; accepted}

\abstract
{}
{
We carried out a project involving the systematic analysis of microlensing data from the Korea 
Microlensing Telescope Network survey. The aim of this project is to identify lensing events 
with complex anomaly features that are difficult to explain using standard binary-lens or 
binary-source models. 
}
{
Our investigation reveals that the light curves of microlensing events KMT-2021-BLG-0284, 
KMT-2022-BLG-2480, and KMT-2024-BLG-0412 display highly complex patterns with three or more 
anomaly features. These features cannot be adequately explained by a binary-lens (2L1S) model 
alone. However, the 2L1S model can effectively describe certain segments of the light curve. 
By incorporating an additional source into the modeling, we identified a 
comprehensive model that accounts for all the observed anomaly features.
}
{
Bayesian analysis, based on constraints provided by lensing observables, indicates that the 
lenses of KMT-2021-BLG-0284 and KMT-2024-BLG-0412 are binary systems composed of M dwarfs. For 
KMT-2022-BLG-2480, the primary lens is an early K-type main-sequence star with an M dwarf 
companion. The lenses of KMT-2021-BLG-0284 and KMT-2024-BLG-0412 are likely located in the bulge, 
whereas the lens of KMT-2022-BLG-2480 is more likely situated in the disk. In all events, the 
binary stars of the sources have similar magnitudes due to a detection bias favoring binary 
source events with a relatively bright secondary source star, which increases detection 
efficiency. 
}
{}

\keywords{gravitational lensing: micro}

\maketitle

\begin{table*}[t]
\caption{Four-body lensing events. \label{table:one}}
\begin{tabular}{lllllll}
\hline\hline
\multicolumn{1}{c}{Event}        &
\multicolumn{1}{c}{Type}         &
\multicolumn{1}{c}{Lens}         &
\multicolumn{1}{c}{Reference}    \\
\hline
OGLE-2006-BLG-109     & 3L1S   &  two planets + host     &  \citet{Gaudi2008}           \\ 
OGLE-2012-BLG-0026    & 3L1S   &  two planets + host     &  \citet{Han2013:0026}        \\
OGLE-2018-BLG-1011    & 3L1S   &  two planets + host     &  \citet{Han2019:1011}        \\     
OGLE-2019-BLG-0468    & 3L1S   &  two planets + host     &  \citet{Han2022:0468}        \\      
KMT-2021-BLG-1077     & 3L1S   &  two planets + host     &  \citet{Han2022:1077}        \\   
KMT-2021-BLG-0240     & 3L1S   &  two planets + host     &  \citet{Han2022:0240}        \\       
OGLE-2006-BLG-284     & 3L1S   &  planet + binary host   &  \citet{Bennett2020}         \\  
OGLE-2007-BLG-349     & 3L1S   &  planet + binary host   &  \citet{Bennett2016}         \\  
OGLE-2008-BLG-092     & 3L1S   &  planet + binary host   &  \citet{Poleski2014}         \\  
OGLE-2016-BLG-0613    & 3L1S   &  planet + binary host   &  \citet{Han2017:0613}        \\  
OGLE-2018-BLG-1700    & 3L1S   &  planet + binary host   &  \citet{Han2020:1700}        \\  
KMT-2020-BLG-0414     & 3L1S   &  planet + binary host   &  \citet{Zang2021a}           \\  
OGLE-2023-BLG-0836    & 3L1S   &  planet + binary host   &  \citet{Han2024:0836}        \\  
KMT-2021-BLG-1122     & 3L1S   &  triple stellar lens    &  \citet{Han2023:1122}        \\    
\hline                                                                                
OGLE-2016-BLG-0882    & 2L2S   &  binary stellar lens    &  \citet{Shin2024}           \\
OGLE-2016-BLG-1003    & 2L2S   &  binary stellar lens    &  \citet{Jung2017}           \\  
KMT-2019-BLG-0797     & 2L2S   &  binary stellar lens    &  \citet{Han2021:0797}       \\  
KMT-2021-BLG-1898     & 2L2S   &  binary stellar lens    &  \citet{Han2022:1898}       \\  
OGLE-2018-BLG-0584    & 2L2S   &  binary stellar lens    &  \citet{Han2023:0584:2119}  \\  
KMT-2018-BLG-2119     & 2L2S   &  binary stellar lens    &  \citet{Han2023:0584:2119}  \\  
MOA-2010-BLG-117      & 2L2S   &  planetary lens         &  \citet{Bennett2018}        \\  
OGLE-2017-BLG-0448    & 2L2S   &  planetary lens         &  \citet{Shin2024}           \\
KMT-2018-BLG-1743     & 2L2S   &  planetary lens         &  \citet{Han2021:1743}       \\  
KMT- 2021-BLG-1547    & 2L2S   &  planetary lens         &  \citet{Han2023:1547}       \\  
\hline                                                                    
OGLE- 2015-BLG-1459   & 1L3S   &  single lens            &  \citet{Hwang2018}          \\  
\hline                                                            
\end{tabular}                                                   
\end{table*}

\section{Introduction} \label{sec:one}

Since the mid-2010s, the Korea Microlensing Telescope Network (KMTNet) team has been conducting 
gravitational microlensing experiments using a network of three wide-field telescopes deployed in 
the Southern Hemisphere \citep{Kim2016}. The data collected from these observations are transmitted 
almost in real-time to the headquarters of the Korea Astronomy and Space Science Institute (KASI) 
for processing. A self-developed algorithm \citep{Kim2018} is employed to identify lensing events, 
and the light curves of these detected events are meticulously examined for any discontinuous 
anomalies. These anomalies undergo careful analysis by multiple researchers to determine their 
origins and confirm whether they are caused by planetary companions to the lens, which is the 
primary objective of the experiment.  Currently, the KMTNet experiment detects over 3,000 
gravitational lensing events annually, with about 10\% of these events exhibiting anomalies of 
various origins \citep{Zang2021b, Zang2022}.  Of these anomalies, approximately 10\% are confirmed 
to be of planetary origin \citep{Gould2022}.

Anomalies in lensing light curves can arise from various factors. The most common cause is the 
binarity of the lens \citep{Mao1991}. In these binary-lens single-source (2L1S) events, caustics 
form on the source plane, and the source's passage through these caustics results in light curves 
that differ from those of single-lens single-source (1L1S) events \citep{Schneider1986}. These 
caustics exhibit complex patterns depending on the separation and mass ratio between the lens 
components and, along with various source trajectories, produce a wide range of anomaly patterns 
\citep{Erdl1993, Han2006, Cassan2008, Gaudi2012}.  Another important cause of anomalies is the 
binarity of the source.  In these single-lens binary-source (1L2S) cases, the event's light curve 
is the superposition of the lensing events occurring for each individual source, which leads to deviations 
in the lensing light curves \citep{Griest1992, Stefano1995, Han1997, Dominik1998, Han1998}.

In some rare cases, the observed anomalies cannot be explained by a three-body lensing model 
(lens plus source) and require a four-body lensing model, which includes an additional lens 
or source component. To date, 14 events have been identified as 3L1S events, in which the lens 
consists of three masses. Of these, six events correspond to planetary systems consisting of a 
host star and two planets, seven involve binary systems that include a planet, and the remaining 
event has been identified as a triple stellar system composed of three stars.  Another type of 
four-body event occurs when both the lens and the source are binaries, known as a 2L2S event. 
So far, ten such events have been identified, four of which involve binary lenses that include 
planetary companions.
In Table~\ref{table:one}, we provide a summary of some of the known four-body lensing 
events, along with a brief description of the corresponding lens systems.

\begin{table*}[t]
\caption{Coordinates, extinction, and baseline magnitude.  \label{table:two}}
\begin{tabular}{lllllll}
\hline\hline
\multicolumn{1}{c}{Event}                      &
\multicolumn{1}{c}{(RA, Dec)$_{\rm J2000}$}    &
\multicolumn{1}{c}{$(l,b)$}                    &
\multicolumn{1}{l}{$A_I$}                      &
\multicolumn{1}{c}{$I_{\rm base}$}             &
\multicolumn{1}{c}{Other ID}             \\
\hline
 KMT-2021-BLG-0284 & (17:58:03.64, -32:18:10.30) & $(-1^\circ\hskip-2pt .5363, -4^\circ\hskip-2pt .0171) $  &  1.48   &  19.49 &  MOA-2021-BLG-072     \\
 KMT-2022-BLG-2480 & (17:35:06.77, -29:54:43.27) & $(-2^\circ\hskip-2pt .0414,  1^\circ\hskip-2pt .4213) $  &  2.71   &  19.68 &                       \\
 KMT-2024-BLG-0412 & (17:55:48.89, -29:54:40.79) & $( 0^\circ\hskip-2pt .2984, -2^\circ\hskip-2pt .4062) $  &  1.54   &  18.79 &  OGLE-2024-BLG-0496   \\
\hline
\end{tabular}
\end{table*}

We conducted a project in which KMTNet data were systematically analyzed to reveal the 
nature of events with complex anomalous features that are challenging to explain. In the 
initial phase, we examined lensing events from the KMTNet survey that exhibited anomalies 
in their light curves, which were then independently analyzed by multiple modelers. Most 
of these anomalies were successfully explained using either 2L1S or 1L2S models. However, 
for a small subset of events in which these three-body models could not account for the 
anomalies, we carried out more detailed analyses using advanced models.  These analyses 
revealed that the difficulty in explaining the anomalies in some events was due to 
significant higher-order effects, such as lens orbital motion, as seen in events like 
OGLE-2018-BLG-0971, MOA-2023-BLG-065, and OGLE-2023-BLG-0136 \citep{Han2024:0971}. In 
other cases, the anomalies were caused by the presence of an additional source or lens 
component, as shown in the 2L2S and 3L1S events summarized in Table~\ref{table:one}.  It 
is important to note that the list of four-body lensing events in the table is not complete, 
as the nature of the anomalies in some events remains unclear, and more advanced models are 
currently being tested to interpret them.  In this work, we present analyses of three 
lensing events with complex anomalies in their light curves that were successfully 
interpreted using 2L2S models.

This paper is organized as follows.  In 
Sect.~\ref{sec:two} we detail the observations conducted to collect the data used for 
our analyses, including a brief description of the instrumentation.  We also outline 
the procedures for data reduction and photometry.  Section~\ref{sec:three} begins with 
the definition of the lensing parameters used in our modeling across different interpretations 
of lensing events.  We then describe the procedure of light curve modeling employed to 
determine these parameters. Subsequent subsections present the analysis process and 
results for each individual event. Each subsection discusses specific anomalies observed 
in the event's light curve, provides model parameters derived from the analysis, and 
outlines the configuration of the lens system.  In Sect.~\ref{sec:four} we identify 
the source stars associated with each event and estimate the angular Einstein radius 
based on the derived information on the source star.  Section~\ref{sec:five} outlines 
the physical quantities of the lens determined based on the observables of individual events. 
Lastly, Sect.~\ref{sec:six} summarizes our findings and presents the conclusions drawn 
from the study.

\section{Observations and data} \label{sec:two}

The three lensing events analyzed in this work are KMT-2021-BLG-0284, KMT-2022-BLG-2480, and 
KMT-2024-BLG-0412.  In Table~\ref{table:two}, we list the equatorial and Galactic coordinates 
of the events, along with the $I$-band extinction ($A_I$) toward the field and the baseline 
magnitude ($I_{\rm base}$).  The anomalous nature of the light curves  for these events was 
identified from the inspection of the data collected by the KMTNet survey. This survey is 
conducted using three identical wide-field telescopes strategically deployed in the Southern 
Hemisphere for continuous observation. Each telescope is located at the Siding Spring Observatory 
in Australia (KMTA), the Cerro Tololo Inter-American Observatory in Chile (KMTC), and the South 
African Astronomical Observatory in South Africa (KMTS). Each KMTNet telescope features a 
1.6-meter aperture, and the mounted camera covers a field of view of 4 square degrees.

Among the analyzed events, KMT-2021-BLG-0284 was additionally observed by the Microlensing
Observations in Astrophysics \citep[MOA;][]{Bond2001, Sumi2003} survey, and KMT-2024-BLG-0412 
was observed by the Optical Gravitational Lensing Experiment \citep[OGLE;][]{Udalski2015}. The 
ID references for these events in the MOA and OGLE surveys are presented in Table~\ref{table:two}. 
The MOA survey operates with a 1.8~m telescope mounted with 2.2 square~degree field camera located 
at Mt. John University Observatory in New Zealand, while the OGLE survey uses a 1.6 m telescope 
equipped with a 1.4 square-degree field camera located at Las Campanas Observatory in Chile. In 
our analysis, we incorporated data obtained from these supplementary observations. Observations 
from the KMTNet and OGLE surveys were primarily conducted in the $I$ band, while the MOA survey 
used a customized MOA-$R$ band with a wavelength range of 609--1109 nm.

The reduction of the data and the photometry of the events were carried out using photometry
pipelines tailored to each survey group: the KMTNet data utilized the \citet{Albrow2009}
pipeline, OGLE data used the \citet{Udalski2003} pipeline, and MOA data were processed with the
\citet{Bond2001} pipeline. To optimize the data, the KMTNet dataset underwent a re-reduction
process using the code developed by \citet{Yang2024}. For each dataset, the error bars from the
photometry pipelines were recalibrated to align with the data scatter and to normalize the $\chi^2$
value per degree of freedom to unity. This normalization process followed the method detailed in
\citet{Yee2012}.

\section{Light curve analyses} \label{sec:three}

The light curve of a lensing event produced by a single mass exhibits a smooth and symmetrical 
shape with respect to the time of the closest lens-source approach \citep{Paczynski1986}.  This 
light curve is characterized by three parameters: the time of closest approach between the lens 
and the source ($t_0$), the separation at that time ($u_0$), and the timescale of the event 
($\te$). The event timescale is defined as the time it takes for the source to traverse the 
Einstein radius ($\thetae$) of the lens, with the impact parameter scaled by $\thetae$.

The departure from the 1L1S magnification pattern in a 2L1S lensing event is primarily caused 
by caustics. Caustics are locations at which the lensing magnification of a point source becomes
infinite. These caustics form closed curves made up of concave segments. The number of these
closed curves varies from one to three, depending on the separation and mass ratio between the
lens components. Representative anomalies in lensing light curves caused by caustics include spike
features occurring as the source crosses a fold of the caustic and bump features as it approaches 
a cusp of the caustic. Caustic spikes manifest as pairs due to the closed structure of the caustic
curve, with the magnification pattern between them typically showing a U-shaped profile.  To 
describe anomalies induced by caustics of a 2L1S event, additional lensing parameters are required.
These parameters are $(s, q, \alpha, \rho)$. The first two parameters represent the projected 
separation and mass ratio between the lens components ($M_1$ and $M_2$). The third parameters 
denotes the angle between the $M_1$--$M_2$ axis and the direction of the source motion relative 
to the lens.  Here, the separation is scaled by $\thetae$.  The last parameter is defined as the 
ratio of the angular source radius ($\theta_*$) to the angular Einstein radius. This parameter 
characterizes the deformation of the lensing light curve by finite source effects during caustic 
crossings or approaches.  This parameter is essential for modeling a subset of 1L1S events with 
very high magnifications, where the lens comes extremely close to the source or crosses over the 
surface of the source star.

In a 2L2S event, which includes a secondary source ($S_2$) alongside the primary source ($S_1$), 
the lensing anomaly can deviate further from that of a 2L1S event.  Consequently, analyzing a 
2L2S event requires additional parameters to characterize the second source: $(t_{0,2}, u_{0,2}, 
q_F)$.  These parameters represent the closest approach time and impact parameter of $S_2$, as 
well as the flux ratio between the two sources.

Based on the caustic spikes commonly observed in the light curves of the analyzed events, 
we initially conducted an analysis based on the 2L1S model.  During this analysis, we 
searched for the binary lens parameters $(s, q)$, for which the lensing magnification 
changes discontinuously with the variation of the parameters, using a grid approach, 
while the remaining parameters, for which lensing magnification smoothly varies with the 
parameter variation, were found using a downhill approach. For the downhill approach, we 
utilized the Markov chain Monte Carlo algorithm.  Subsequently, we refined the model 
by allowing all parameters to vary.

For the analyzed events, all of which displayed highly complex light curves with three or more
anomaly features, it was found that the lensing light curves could not be adequately described 
by a 2L1S model alone.  Subsequently, we investigated whether the 2L1S model could sufficiently 
describe segments of the light curve while excluding specific anomaly features. If this approach 
was feasible, we then considered a 2L2S model by introducing an additional source to determine 
if the second source could account for the omitted segments of the light curve.  We also 
considered the possibility of explaining the unexplained portion by introducing an additional 
lens component. However, 3L1S models failed to account for the anomalies in all of the events 
analyzed.

\subsection{KMT-2021-BLG-0284} \label{sec:three-one}

\begin{figure}[t]
\includegraphics[width=\columnwidth]{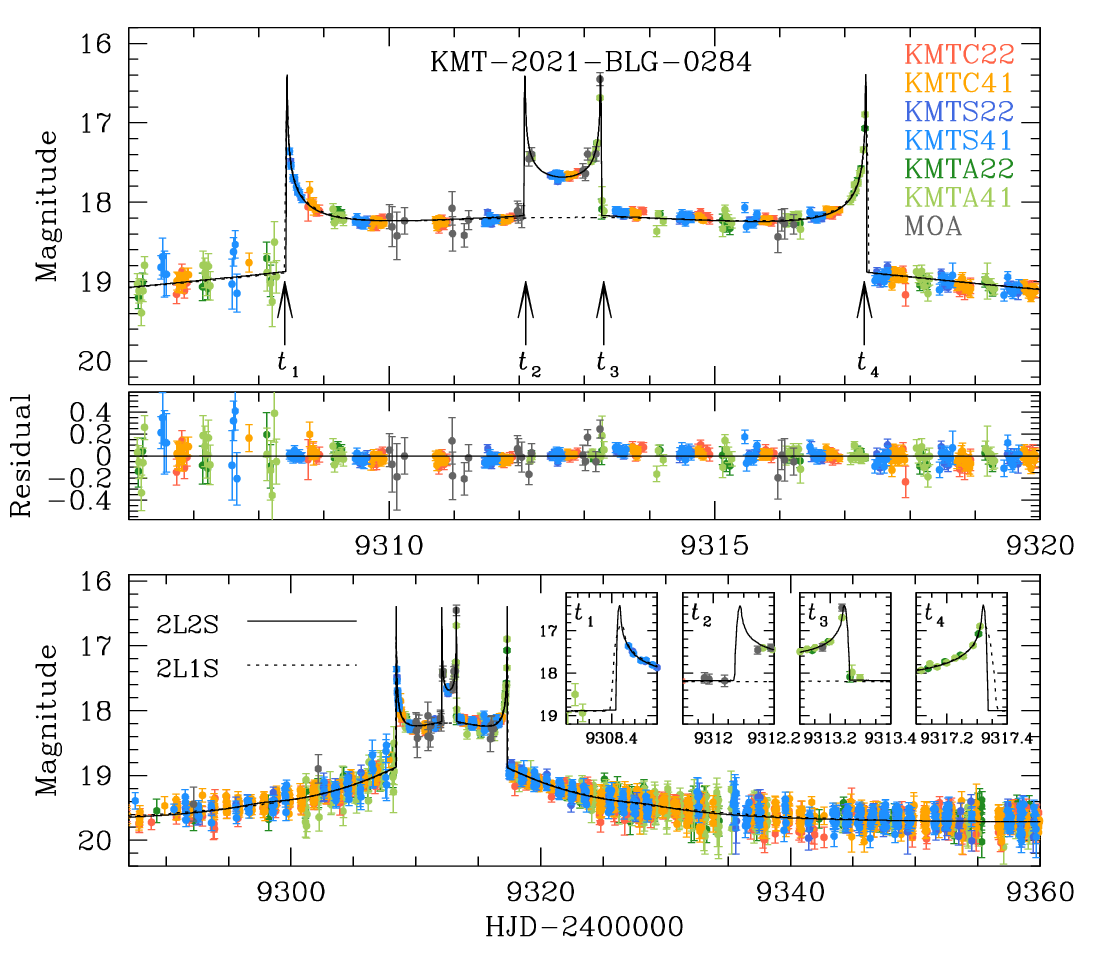}
\caption{
Lensing light curve of KMT-2021-BLG-0284.  The lower panel provides an overall view, while the
upper panel offers a zoomed-in perspective of the anomaly region. Arrows labeled $t_1$, $t_2$, 
$t_3$, and $t_4$ indicate the times of the caustic-crossing features. Insets within the lower 
panel present an enlarged view around each individual caustic peak. The colors of the data points 
correspond to the telescopes and observation fields, as indicated in the legend. The solid 
curve overlaying the data points represents the best-fit 2L2S model, while the dotted curve 
depicts a 2L1S model derived by fitting the data, excluding those around $t_2$ and $t_3$.
}
\label{fig:one}
\end{figure}

The lensing event KMT-2021-BLG-0284 was initially detected by the KMTNet survey on April 6, 
2021, two days after the first caustic peak. This date corresponds to the abridged heliocentric
Julian date ${\rm HJD}^\prime \equiv {\rm HJD}-2450000=9310$.  Figure~\ref{fig:one} 
presents the light curve of the event. It features an intricate magnification pattern with four 
caustic spikes occurring at approximately ${\rm HJD}^\prime =9308.4$ ($t_1$), 9312.1 ($t_2$), 
9313.3 ($t_3$), and 9317.3 ($t_4$).  The source was situated in the narrow strip region for 
which the KMTNet BLG22 and BLG41 fields overlap. The observation cadence was 1.0 hour for the 
BLG22 field and 0.5 hour for the BLG41 field. One day after the KMTNet announcement of event 
detection, the MOA group also reported the detection of the event with a reference ID 
MOA-2021-BLG-072. As the event progressed, it exhibited three additional caustic-crossing 
features.

\begin{figure}[t]
\includegraphics[width=\columnwidth]{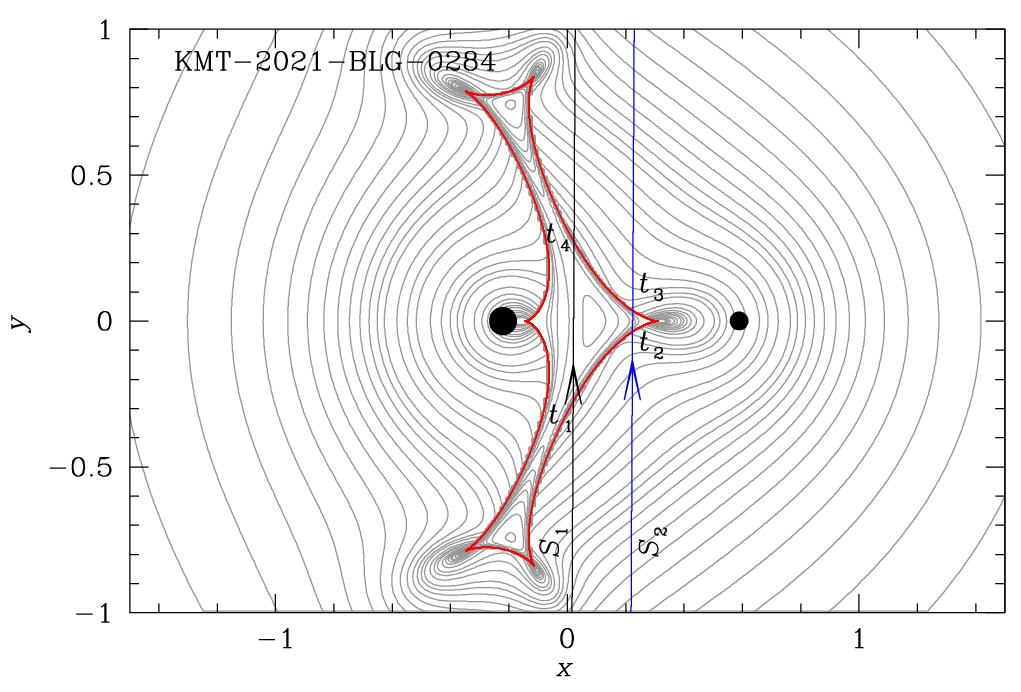}
\caption{
Lens system configuration of KMT-2021-BLG-0284. The red figures composed of concave curves 
represents the caustic. The two black filled dots indicate the positions of the lens components; the bigger one represents the heavier component. The black and blue arrowed lines 
represent the trajectories of the primary ($S_1$) and secondary ($S_2$) source stars, respectively.  
The arrow tips on the trajectories are shown at the same moment in time.  The points marked as 
$t_1$, $t_2$, $t_3$, and $t_4$ on the source trajectories represent the source positions at the 
times of the anomalies marked in Fig.~\ref{fig:one}.  Gray curves encompassing the caustic are 
equi-magnification contours.    
}
\label{fig:two}
\end{figure}

\begin{table}[h]
\caption{Best-fit parameters of KMT-2021-BLG-0284.\label{table:three}}
\begin{tabular*}{\columnwidth}{@{\extracolsep{\fill}}lllcc}
\hline\hline
\multicolumn{1}{c}{Parameter}    &
\multicolumn{1}{c}{Value}        \\
\hline
  $t_{0,1}$ (HJD$^\prime$)   &   $ 9312.917 \pm 0.062 $ \\   
  $u_{0,1}$                  &   $ -0.0217 \pm 0.00329$ \\   
  $t_{0,2}$ (HJD$^\prime$)   &   $ 9312.650 \pm 0.027 $ \\    
  $u_{0,2}$                  &   $ -0.2240 \pm 0.0051 $ \\      
  $\te$ (days)               &   $ 16.25 \pm 0.29     $ \\       
  $s$                        &   $ 0.8095 \pm 0.0080  $ \\     
  $q$                        &   $ 0.3742 \pm 0.0096  $ \\   
  $\alpha$ (rad)             &   $ 1.5762 \pm 0.0083  $ \\      
  $\rho_1$ (10$^{-3}$)       &   $ < 0.1              $ \\
  $\rho_2$ (10$^{-3}$)       &   --                     \\
  $q_F$                      &   $ 0.766 \pm 0.01     $ \\
\hline                                                   
\end{tabular*}
\tablefoot{ ${\rm HJD}^\prime = {\rm HJD}- 2450000$.}
\end{table}

The anomaly pattern in the light curve was challenging to interpret using a 2L1S model.  Based on
its appearance, the first and fourth features (at $t_1$ and $t_4$) seem to form a pair of caustic 
spikes caused by the source star entering and exiting a caustic. Similarly, the second and third 
features (at $t_2$ and $t_3$) appear to be another pair of caustic spikes. The presence of this 
second pair in the region between the features of the first caustic pair strongly suggests a 
departure of the lens system from a 2L1S configuration. Based on this, Yuki Hirao of the MOA group 
released a triple lens (3L1S) model on April 16, 2021.  Although this model roughly outlined the 
overall anomaly pattern, it left noticeable residuals. On April 21, 2021, Cheongho Han proposed 
another interpretation based on a 2L2S model.  Upon reanalysis using re-reduced data conducted 
after the event concluded, it was found that the 2L2S model significantly better explained the 
observed data than the 3L1S model.

The model curve of the 2L2S solution and its residual are displayed in Fig.~\ref{fig:one}. The 
estimated binary lens parameters are $(s, q)\sim (0.81, 0.37)$, and the event timescale is $\te 
\sim 16.3$~days. The complete lensing parameters of the solution are listed in Table~\ref{table:three}.  
As illustrated in the four insets of the lower panel, none of the four caustic crossings were 
sufficiently resolved to reliably measure the normalized source radii for the first source ($\rho_1$) 
or the second source ($\rho_2$), with only a loose upper limit for $\rho_1$ being constrained.  
The estimated flux ratio between the primary and secondary source stars is $q_F\sim 0.77$.

Figure~\ref{fig:two} illustrates the lens system configuration, depicting the source trajectories 
of $S_1$ and $S_2$ relative to the positions of the lens and caustic.  The solution indicates that 
the lens creates a resonant caustic, elongated in a direction perpendicular to the $M_1$--$M_2$ 
axis. The primary source passed through the region between the lens components, crossing the caustic 
at $t_1$ and $t_4$.  Meanwhile, the secondary source approached the caustic with a larger impact 
parameter, crossing the tip of the right-side caustic at $t_2$ and $t_3$.

\begin{figure}[t]
\includegraphics[width=\columnwidth]{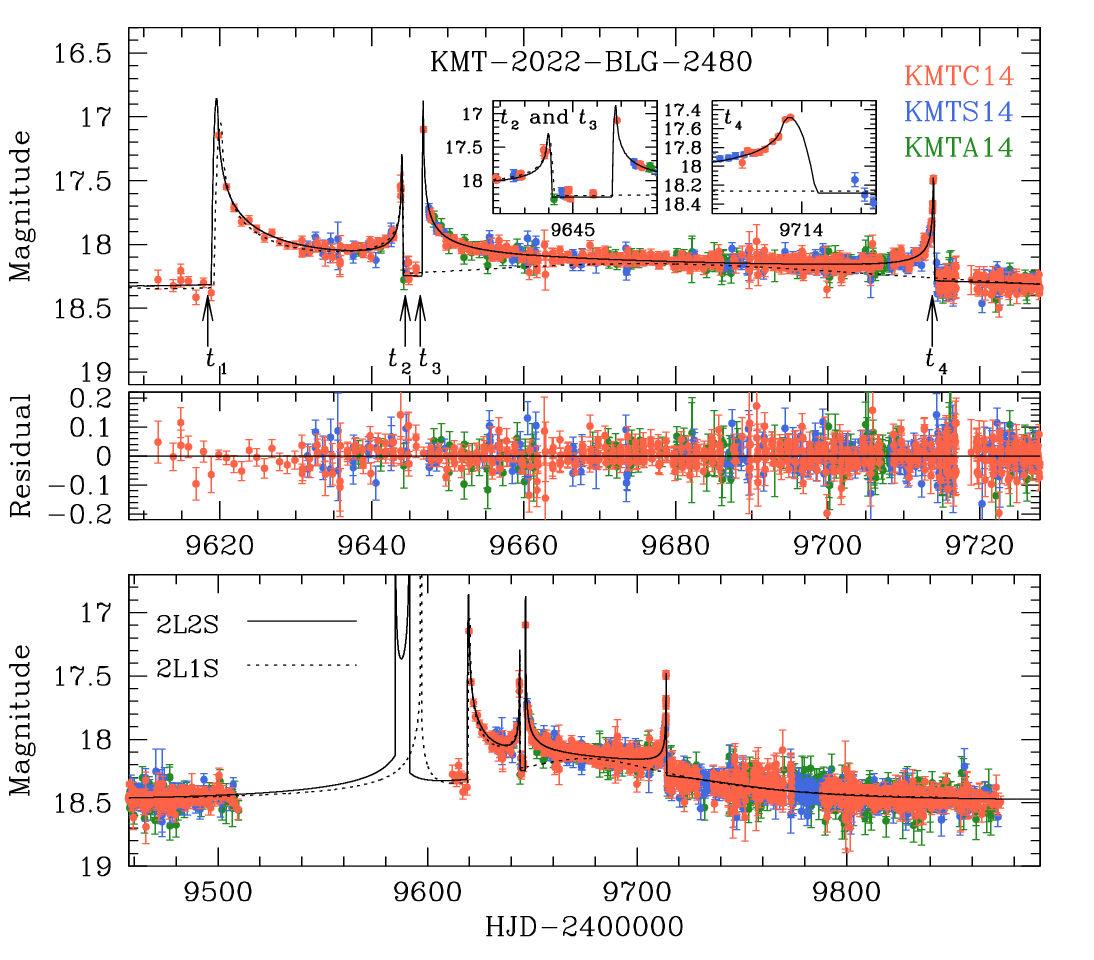}
\caption{
Light curve of the lensing event KMT-2022-BLG-2480. The notations are the same as those in 
Fig.~\ref{fig:one}. The insets in the upper panel show enlargements of the regions around 
$t_2$, $t_3$, and $t_4$.
}
\label{fig:three}
\end{figure}

\subsection{KMT-2022-BLG-2480} \label{sec:three-two}

The lensing event KMT-2022-BLG-2480 was detected through the EventFinder algorithm \citep{Kim2018} 
and observed exclusively by the KMTNet group.  The lensing-induced magnification of the source 
flux started before the 2022 season and continued through to the end of the 2024 season.  The 
source was located in the KMTNet BLG14 field, which was observed with an hourly cadence.  
Figure~\ref{fig:three} displays the lensing light curve of event. Similar to that of 
KMT-2021-BLG-0284, it displays complex magnification pattern characterized by four caustic 
spikes occurring at around ${\rm HJD}^\prime = 9618.5$ ($t_1$), 9644.4 ($t_2$), 9646.6 ($t_3$), 
and 9713.8 ($t_4$). The difference from the event KMT-2021-BLG-0284 is that the two pairs of 
caustic-crossing features are not overlapping but separated.

Upon analyzing the light curve, we found that the anomaly pattern of KMT-2022-BLG-2480 could
not be adequately described by a 2L1S configuration.  In order to assess whether a 2L1S model 
could fit a portion of the light curve, we conducted an additional modeling excluding the data 
between $t_3$ and $t_4$. This analysis yielded a model that explains the anomaly patterns at 
$t_1$ and $t_2$.  Figure~\ref{fig:three} illustrates the model curve of this 2L1S solution. Subsequently, we conducted 2L2S modeling to further elucidate the other caustic 
features at $t_3$ and $t_4$. Through this, we identified a model that comprehensively describes 
all observed anomaly features. The 2L2S model found from the analysis and its residual are 
presented in Fig.~\ref{fig:three}.  According to this model, there is another set of caustic 
features located around ${\rm HJD}^\prime \sim 9685$, which falls before the observation period 
and therefore was not observed.

Table~\ref{table:four} lists the lensing parameters of the 2L2S solution.  The estimated binary-lens 
parameters are $(s, q) \sim (1.1, 0.35)$.  The event timescale, $\te \sim 107$~days, is significantly 
longer than the typical 10--20 day timescales of lensing events produced by low-mass stars \citep{Han2003}.
Among the four caustics, only the last caustic is partially resolved. The first two spikes at $t_1$ and 
$t_2$ were produced by the caustic crossings of the primary source, and thus its normalized source 
radius could not be constrained.  The normalized radius of the second source, derived from the 
partially resolved last peak, is $\rho_2 \sim 0.9 \times 10^{-3}$.  The flux from the source generating 
the second set of caustic spikes ($S_2$) is estimated to be 0.68 times less than the flux from the source 
producing the first set of spikes ($S_1$).  In events with a long timescale, signals caused by parallax 
effects, which arise from the deviation of the observer's motion from rectilinear due to Earth's orbit 
\citep{Gould1992}, can often be detected in the light curve. However, in the case of the 
KMT-2022-BLG-2480 event, it was difficult to clearly detect the parallax effect due to significant 
photometric errors caused by the faintness of the source stars.

\begin{figure}[t]
\includegraphics[width=\columnwidth]{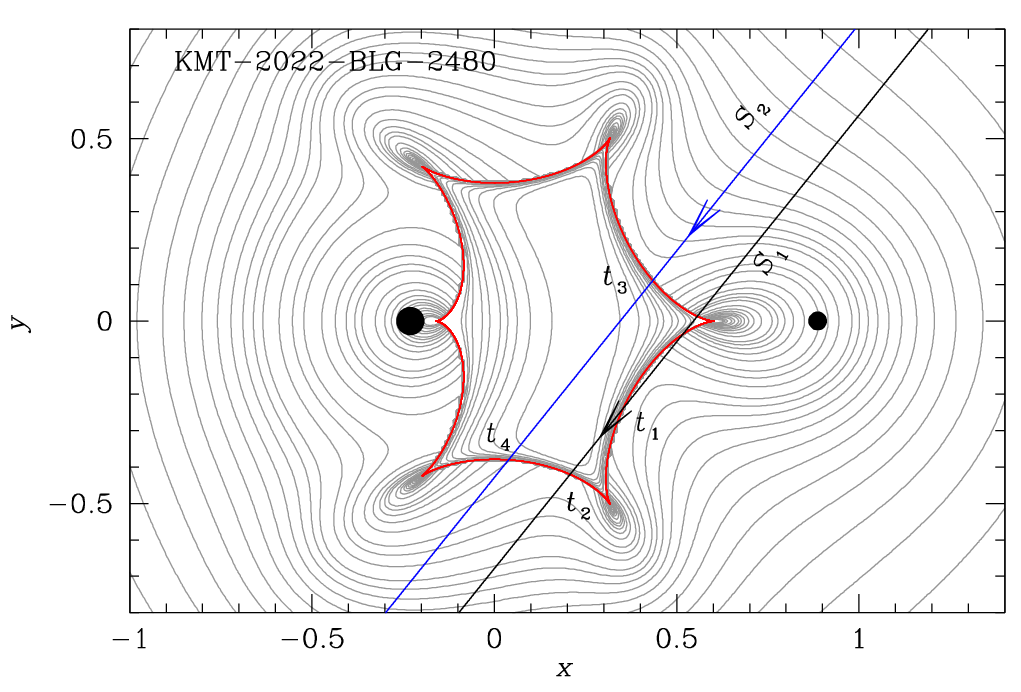}
\caption{
Configuration of the KMT-2022-BLG-2480 lens system. Notations are the same as those in Fig.~\ref{fig:two}.
}
\label{fig:four}
\end{figure}

\begin{table}[t]
\caption{Best-fit parameters of KMT-2022-BLG-2480.\label{table:four}}
\begin{tabular*}{\columnwidth}{@{\extracolsep{\fill}}lllcc}
\hline\hline
\multicolumn{1}{c}{Parameter}    &
\multicolumn{1}{c}{Value}        \\
\hline
  $t_{0,1}$ (HJD$^\prime$)   &   $ 9623.007 \pm 0.856$ \\   
  $u_{0,1}$                  &   $ 0.4254 \pm 0.0046 $ \\   
  $t_{0,2}$ (HJD$^\prime$)   &   $ 9685.002 \pm 0.476$ \\    
  $u_{0,2}$                  &   $ 0.2689 \pm 0.0027 $ \\      
  $\te$ (days)               &   $ 106.64 \pm 0.74   $ \\       
  $s$                        &   $ 1.1173 \pm 0.0043 $ \\     
  $q$                        &   $ 0.346 \pm 0.011   $ \\   
  $\alpha$ (rad)             &   $ -0.893 \pm 0.013  $ \\      
  $\rho_1$ (10$^{-3}$)       &   --                    \\
  $\rho_2$ (10$^{-3}$)       &   $ 0.89 \pm 0.06     $ \\
  $q_F$                      &   $ 0.682 \pm 0.019   $ \\
\hline                                                   
\end{tabular*}
\end{table}

Figure~\ref{fig:four} shows the configuration of the KMT-2022-BLG-2480 lens system.  The binary lens 
forms a resonant caustic with six folds. The primary source crossed the caustic four times: it initially 
entered and exited the caustic by passing through the right on-axis cusp, then reentered and exited 
again through the lower right cusp. The first two crossings produced an unobserved pair of caustic 
spikes, while the latter two crossings resulted in spikes at $t_1$ and $t_2$. The second source 
diagonally passed through the caustic, producing a spike at $t_3$ upon entry and another at $t_4$ 
upon exit.

\subsection{KMT-2024-BLG-0412} \label{sec:three-three}

\begin{figure}[t]
\includegraphics[width=\columnwidth]{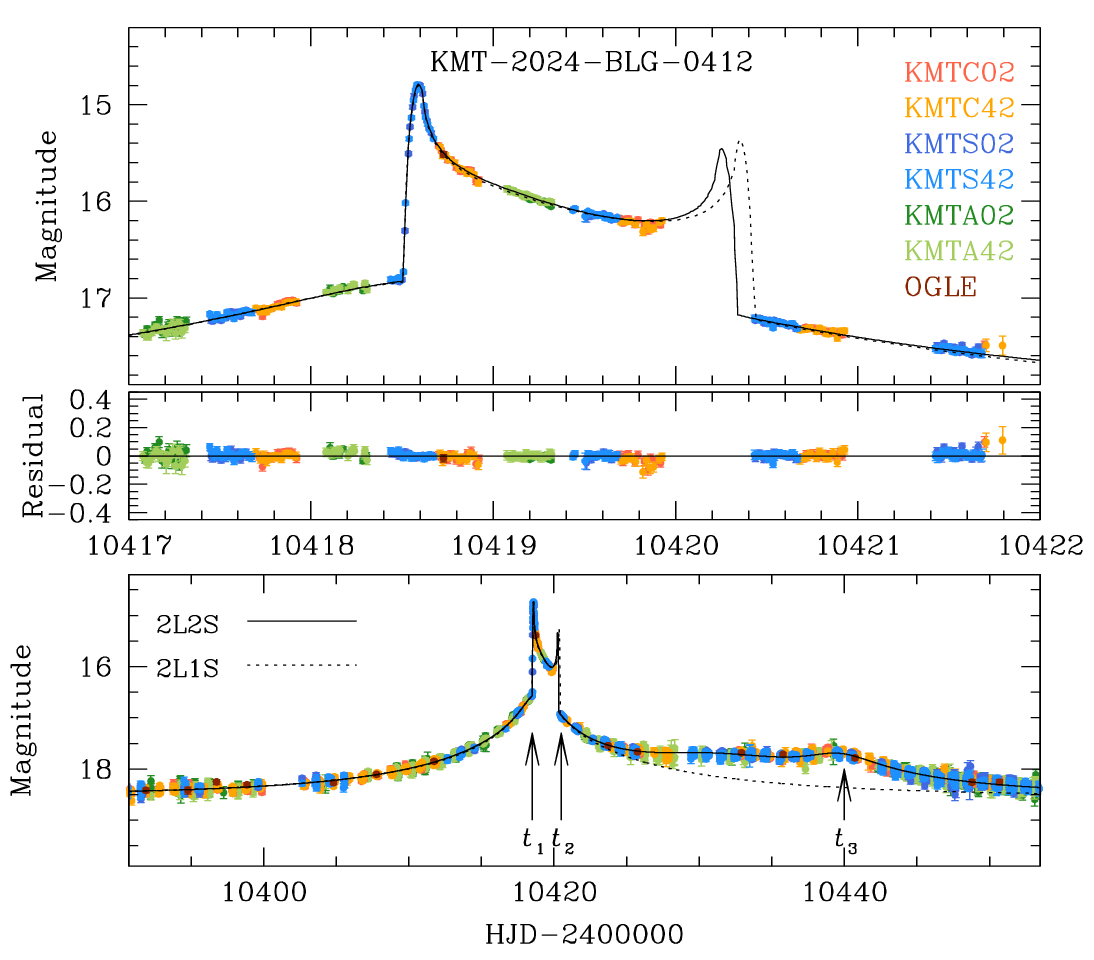}
\caption{
Light curve of the lensing event KMT-2024-BLG-0412.     
}
\label{fig:five}
\end{figure}

The lensing event KMT-2024-BLG-0412 was first detected by
the KMTNet group on April 5, 2024 (${\rm HJD}^\prime =10405$) during its rising stage. The source 
lay in the overlapping region of the KMTNet prime fields BLG02 and BLG42, each observed with a 
0.5-hour cadence, resulting in a combined cadence of 0.25 hours. The event was later identified 
by the OGLE survey, which observed it with a relatively low cadence. The OGLE ID reference of 
the event is OGLE-2024-BLG-0496.  The light curve of the event is displayed in Fig.~\ref{fig:five}.  
On ${\rm HJD}^\prime =10418.5$ ($t_1$), the event's light curve exhibited a sharp rise due to a 
caustic crossing. Based on the U-shaped pattern, another spike was expected around $t_2=10420.5$, 
although this was not observed due to bad weather at the Australian site. After the caustic exit, 
the light curve did not descend but remained steady, then reached another weak peak around 
${\rm HJD}^\prime =10440$ ($t_3$) before returning to the baseline.

As with the two previous events, the light curve of the KMT-2024-BLG-0412 event was difficult to
explain with a 2L1S model.  However, modeling conducted with data excluding the region around the 
$t_3$ bump revealed that the caustic feature could be well described by a 2L1S model. The model 
curve of this 2L1S solution is shown in Fig.~\ref{fig:five}. From a subsequent 
modeling based on the 2L2S interpretation, we found a solution that explains all the anomaly 
features in the lensing light curve. A 3L1S interpretation does not yield a model with a fit 
equivalent to that of the 2L2S solution

Table~\ref{table:five} presents the lensing parameters obtained from the 2L2S modeling.  We 
identified a pair of solutions resulting from the close--wide degeneracy.  The parameters 
describing the binary lens are $(s, q)_{\rm close}\sim (0.56, 0.26)$ for the close solution 
and $(s, q)_{\rm wide}\sim (2.79, 0.62)$ for the wide solution.  It was found that the wide 
solution provides a significantly better fit than the close solution, with $\Delta\chi^2 = 
351.4$.  Therefore, subsequent analysis is based on the wide solution.  The event timescale 
was measured as $\te \sim 22$~days.

\begin{figure}[t]
\includegraphics[width=\columnwidth]{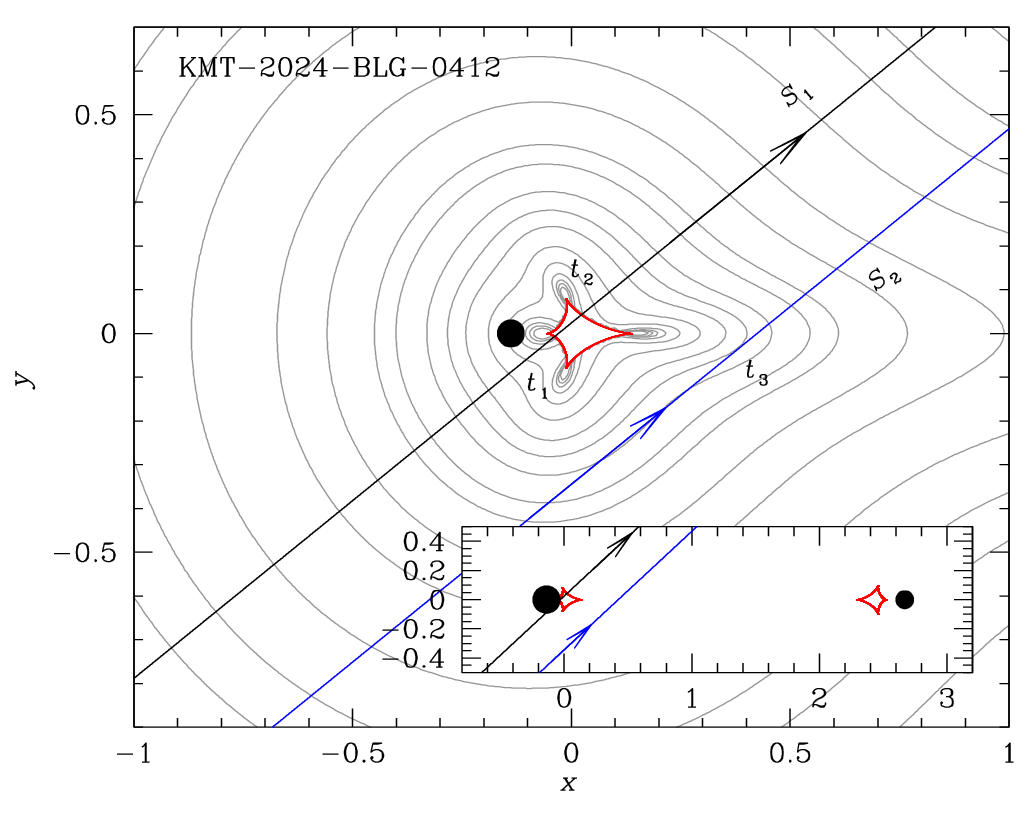}
\caption{
Configuration of the lens system
KMT-2024-BLG-0412. The inset provides an overall
view, while the main panel presents a zoomed-in
view of the area surrounding the heavier lens
component.      
}
\label{fig:six}
\end{figure}

The lens system configuration of KMT-2024-BLG-0412 is shown in Fig.~\ref{fig:six}.  The binary lens 
system produced two sets of caustics: one located near the heavier lens component ($M_1$) and another 
near the other lens component ($M_2$). The primary source first approached the lens, followed by the 
secondary source trailing behind. Caustic spikes at $t_1$ and $t_2$ were generated by the primary source 
star's crossings over the caustic lying near $M_1$, while the bump around $t_3$ was produced by the 
approach of the secondary source to the protruding cusp of the caustic.  Consequently, the normalized 
source radius of $S_1$ was measured, whereas that of $S_2$ could not be constrained. The estimated flux 
ratio $q_F \sim 1.1$ suggests that the source star crossing the caustic ($S_1$) is slightly fainter than 
the source star that does not cross the caustic ($S_2$).

\begin{table}[t]
\caption{Best-fit parameters of KMT-2024-BLG-0412.\label{table:five}}
\begin{tabular*}{\columnwidth}{@{\extracolsep{\fill}}lllcc}
\hline\hline
\multicolumn{1}{c}{Parameter}    &
\multicolumn{1}{c}{Close}        &
\multicolumn{1}{c}{Wide}         \\
\hline
  $\chi^2$                   &  $4877.6             $  &  $4526.2             $  \\ 
  $t_{0,1}$ (HJD$^\prime$)   &  $10419.210 \pm 0.004$  &  $10419.307 \pm 0.011$  \\   
  $u_{0,1}$                  &  $2.240 \pm 0.065    $  &  $1.872 \pm 0.029    $  \\   
  $t_{0,2}$ (HJD$^\prime$)   &  $10436.513 \pm 0.038$  &  $10433.324 \pm 0.093$  \\    
  $u_{0,2}$                  &  $-0.2852 \pm 0.0035 $  &  $-0.2705 \pm 0.0059 $  \\      
  $\te$ (days)               &  $16.34 \pm 0.07     $  &  $21.80 \pm 0.22     $  \\       
  $s$                        &  $0.563 \pm 0.001    $  &  $2.794 \pm 0.016    $  \\     
  $q$                        &  $0.262 \pm 0.003    $  &  $0.624 \pm 0.024    $  \\   
  $\alpha$ (rad)             &  $2.3531 \pm 0.0033  $  &  $2.4579 \pm 0.0046  $  \\      
  $\rho_1$ (10$^{-3}$)       &  $3.36 \pm 0.02      $  &  $2.43 \pm 0.03      $  \\
  $\rho_2$ (10$^{-3}$)       &  --                     &  --                     \\
  $q_F$                      &  $0.814 \pm 0.013    $  &  $1.118 \pm 0.025    $  \\
\hline                                                   
\end{tabular*}
\end{table}

\begin{figure}[t]
\centering
\includegraphics[width=\columnwidth]{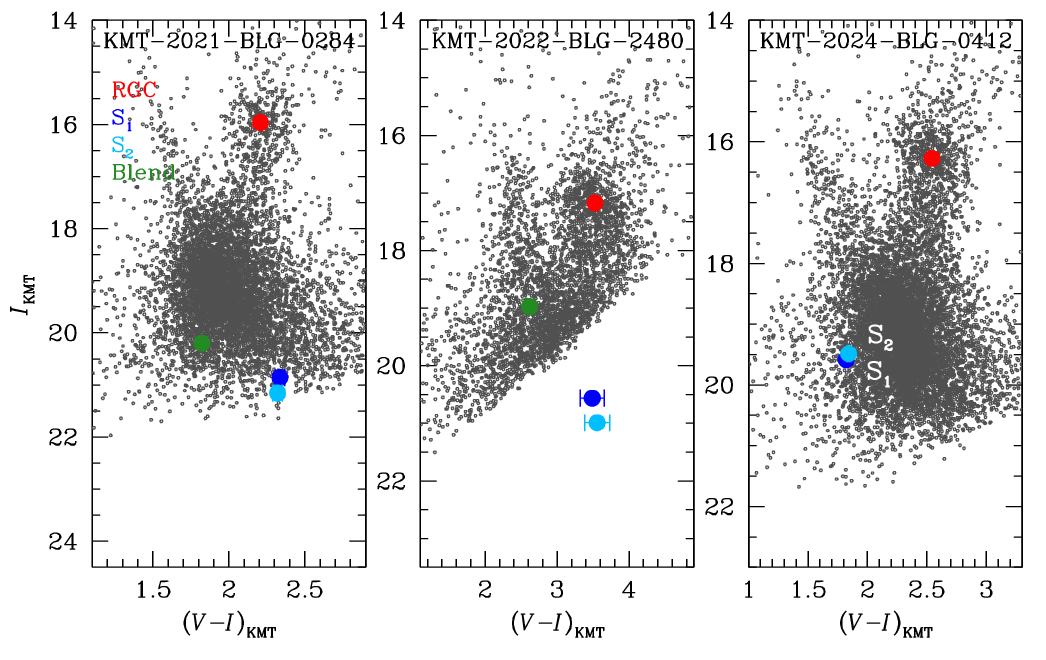}
\caption{
Locations of source stars (blue dot for the primary source and cyan dot for the secondary source) and the
centroid of red giant clumps (RGC, red dot) for the lensing events KMT-2021-BLG-0284, KMT-2022-BLG-2480, 
and KMT-2024-BLG-0412. For KMT-2021-BLG-0284 and KMT-2022-BLG-2480, we also mark the location of the
blend (green dot).      
}
\label{fig:seven}
\end{figure}

\begin{table*}[t]
\caption{Source parameters, angular source radii, Einstein radii, and relative lens-source proper motions.\label{table:six}}
\begin{tabular}{lllll}
\hline\hline
\multicolumn{1}{c}{Parameter}           &
\multicolumn{1}{c}{KMT-2021-BLG-0284}   &
\multicolumn{1}{c}{KMT-2022-BLG-2480}   &
\multicolumn{1}{c}{KMT-2024-BLG-0412}   \\
\hline
$(V-I, I)_{S_1}$            &  $(2.336 \pm 0.031, 20.859 \pm 0.009)  $  &  $(3.489 \pm 0.165, 20.563 \pm 0.013)  $  &  $(1.889 \pm 0.020, 19.582 \pm 0.011) $  \\
$(V-I, I)_{S_2}$            &  $(2.321 \pm 0.032, 21.156 \pm 0.009)  $  &  $(3.556 \pm 0.174, 20.991 \pm 0.017)  $  &  $(1.910 \pm 0.019,19.486 \pm 0.011)  $  \\
$(V-I, I)_{\rm RGC}$        &  $(2.208, 15.965)                      $  &  $(3.527, 17.175)                      $  &  $(2.544, 16.275)                     $  \\
$(V-I, I)_{{\rm RGC},0}$    &  $(1.060, 14.501)                      $  &  $(1.060, 14.526)                      $  &  $(1.060, 14.417)                     $  \\
$(V-I,I)_{S_1,0}$           &  $(1.188 \pm 0.031, 19.395 \pm 0.009)  $  &  $(1.022 \pm 0.165, 17.914 \pm 0.013)  $  &  $(0.405 \pm 0.020, 17.723 \pm 0.011) $  \\
$(V-I, I)_{S_2,0}$          &  $(1.173 \pm 0.032, 19.692 \pm 0.009)  $  &  $(1.089 \pm 0.174, 18.341 \pm 0.017)  $  &  ($0.426 \pm 0.019, 17.628 \pm 0.0110 $  \\
$(V-I, I)_b$                &  $(1.822, 20.200)                      $  &  $(2.618, 18.973)                      $  &   --                                     \\
 Type ($S_1$)               &   K4V                                     &   K3IV                                    &   F2V                                    \\
 Type ($S_2$)               &   K4V                                     &   K3IV                                    &   F2V                                    \\
 $\theta_{*,S_1}$ ($\mu$as) &  $0.719 \pm 0.055                      $  &  $1.170 \pm 0.209                      $  &  $0.660 \pm 0.048                     $  \\
 $\theta_{*,S_2}$ ($\mu$as) &  $0.613 \pm 0.047                      $  &  $1.03 \pm 0.19                        $  &  $0.702 \pm 0.051                     $  \\
 $\thetae$ (mas)            &   --                                      &  $0.98 \pm 0.19                        $  &  $0.272 \pm 0.020                     $  \\
 $\mu$ (mas/yr)             &   --                                      &  $3.35 \pm 0.66                        $  &  $4.55 \pm 0.33                       $  \\
\hline
\end{tabular}
\end{table*}

\section{Source stars and angular Einstein radii} \label{sec:four} 

In this section we define the stars that make up the binary source. Accurately defining the source 
is crucial not only for comprehensively characterizing the events but also for measuring the angular 
Einstein radii of the events. The angular Einstein radius is estimated by combining the measured 
normalized source radius with the angular source radius $\theta_*$, using the relation 
\begin{equation}
\thetae = {\theta_* \over \rho}.
\label{eq1}
\end{equation}
We defined the source star for each event by measuring its color and magnitude. To achieve this, 
we first determined the combined source flux from the primary and secondary source stars. This 
was done by regressing the photometric data ($F_{\rm obs}$) with respect to model 
$A_{\rm model}(t)$:
\begin{equation}
F_{\rm obs}(t)= F_S A_{\rm model}(t) + F_b.
\label{eq2}
\end{equation}
Here, $F_{S}$ represents the combined flux from the primary ($F_{S_1}$) and secondary ($F_{S_2}$) 
source stars, and $F_b$ represents the flux from a blend. Using the measured flux ratio $q_F$ 
between the source stars, the flux from each individual source star was estimated as
\begin{equation}
F_{S_1} =\left( {1 \over 1+q_F} \right)F_S; \qquad
F_{S_2} =\left({q_F \over 1+q_F}\right)F_S.
\label{eq3}
\end{equation}
We measured the flux values in the two passbands of $I$ and $V$ for the estimation of the source 
color. This process was done with the use of the photometric data processed using the pyDIA code 
\citet{Albrow2017}.

Figure~\ref{fig:seven} illustrates the source positions for the events on the instrumental 
color-magnitude diagrams (CMDs) of stars near the source. These CMDs were created using KMTC datasets, and the photometry of the stars was performed with the same pyDIA code used for measuring the 
source flux.  For KMT-2021-BLG-0284 and KMT-2022-BLG-2480, we identified the positions of the blend.  Table~\ref{table:six} lists the colors and magnitudes of the 
primary source star, $(V-I, I)_{S_1}$, secondary star, $(V-I, I)_{S_2}$, and blend, $(V-I, I)_b$. 
In all events, the binary stars that make up the source are found to have similar magnitudes, with 
a magnitude difference of less than 0.5~mag.  \citet{Han1998} noted that binary-source events are 
detected less frequently than binary-lens events because the effect of the secondary star's flux in 
a binary source event is relatively minor compared to the mass effect in a binary lens event.  The 
similarity in brightness of the stars in the binary source for the analyzed lensing events is likely 
due to the higher detection efficiency for binary source events with a relatively bright secondary 
source star.  We observe that the source stars of KMT-2024-BLG-0412 are substantially bluer than 
typical bulge main-sequence stars, suggesting they may be located in the far disk behind the bulge.

\begin{figure}[t]
\includegraphics[width=\columnwidth]{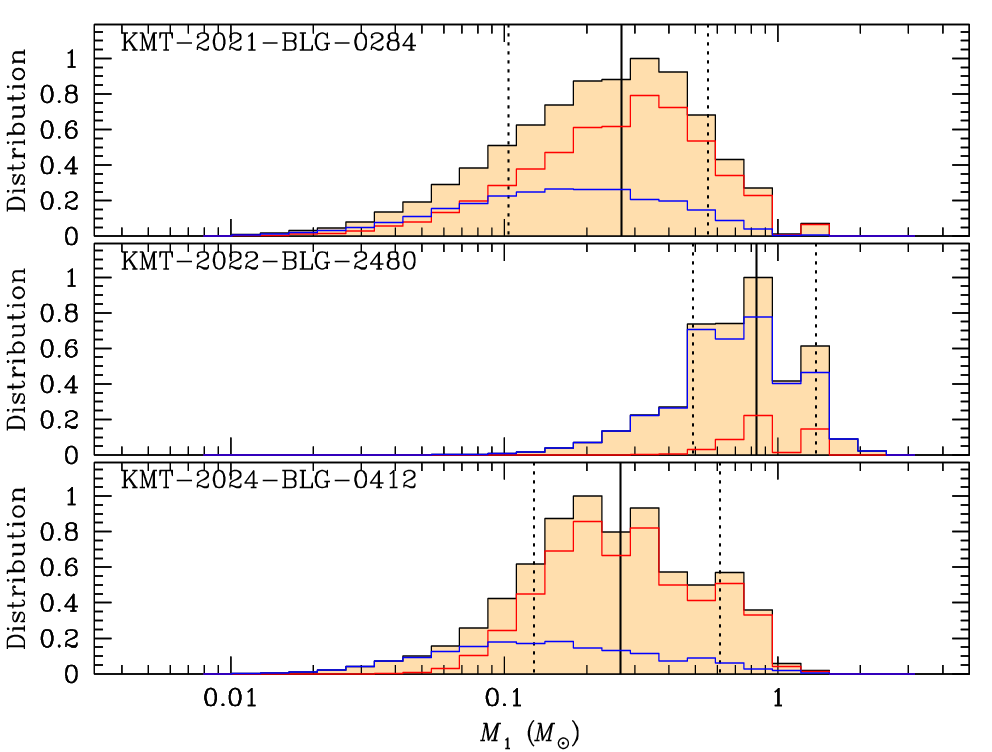}
\caption{
Bayesian posteriors of the mass for the heavier lens component. In each posterior, the red and 
blue curves illustrate the contributions from the disk and bulge lens populations, respectively, 
while the black curve represents the combined contributions from the two populations. The solid 
vertical line marks the median, and the two dotted lines denote the 1$\sigma$ range of the 
distribution.   
}
\label{fig:eight}
\end{figure}

The measured instrumental color and magnitude of the source, $(V-I, I)_S$, were calibrated using 
the method of \citet{Yoo2004}.  This method utilizes the centroid of the red clump giant (RGC), 
$(V-I, I)_{\rm RGC}$, in the CMD as a reference for calibration. By measuring the offset between the 
source and the RGC centroid in the CMD, $\Delta(V-I, I)$, and using the known de-reddened color and 
magnitude of the RGC centroid, $(V-I, I)_{{\rm RGC},0}$, the de-reddened color and magnitude of the 
source are calibrated as 
\begin{equation}
(V-I, I)_{S,0} = (V-I, I)_{{\rm RGC},0} + \Delta(V-I, I).
\label{eq4}
\end{equation}
We adopted the $(V-I, I)_{{\rm RGC},0}$ values from \citet{Bensby2013} and \citet{Nataf2013}. 
The positions of the RGC centroids in the CMDs for each event are indicated in 
Fig.~\ref{fig:seven}, with their values detailed in Table~\ref{table:six}. This table also 
presents the estimated de-reddened colors and magnitudes of the binary source stars for each 
event: $(V-I, I)_{S_1,0}$ for $S_1$ and $(V-I, I)_{S_2,0}$ for $S_2$.

The angular radius of the source is determined based on its measured color and magnitude. 
For this determination, $V-I$ was initially converted to $V-K$ using the color-color relation
established by \citet{Bessell1988}. Subsequently, the angular source radius was derived from
the relation between $(V-K, V)$ and $\theta_*$ provided by \citet{Kervella2004}. Using this 
angular radius, the Einstein radius is calculated using Eq.~(\ref{eq1}). Additionally, in 
conjunction with the event timescale, the relative proper motion between the lens and source 
is estimated as $\mu  = \thetae / \te$.
Table~\ref{table:six} presents the estimated values of $\theta_*$, $\thetae$, and $\mu$ for 
each individual event. For KMT-2021-BLG-0284, the angular Einstein radius could not be determined 
because the normalized source radius was not measured for either the primary or secondary source 
star. For KMT-2022-BLG-2480, where the normalized source radius was measured only for the secondary 
source, the angular Einstein radius was estimated as $\thetae = \theta_{*,S_2}/\rho_2$.  In contrast, 
for KMT-2024-BLG-0496, the normalized source radius was measured only for the primary source, and 
the angular Einstein radius was estimated as $\thetae = \theta_{*,S_1}/\rho_1$.

\section{Physical lens parameters} \label{sec:five} 

In this section we determine the physical lens parameters, including the mass ($M$) and distance 
($\dl$) to the binary lens systems.  The lens parameters are constrained by the lensing observables 
$\te$, $\thetae$, and $\pie$, where $\pie$ represents the microlens parallax.  These observables 
are related to the lens parameters as
\begin{equation}
\te = {\thetae\over\mu}; \qquad
\thetae = (\kappa M \pi_{\rm rel})^{1/2};\qquad
\pie = {\pi_{\rm rel} \over \thetae}, 
\label{eq5}
\end{equation}
where $\kappa = 4G/(c^2{\rm AU}) \simeq 8.14~{\rm mas}/M_\odot$, 
$\pi_{\rm rel}={\rm AU}(D_{\rm L}^{-1} - D_{\rm S}^{-1})$ 
represents the relative lens-source parallax, and $\ds$ denotes the distance to the source. 
Of the observables, the event timescale is measured for all events, while the 
microlens parallax is measured for none. The angular Einstein radius is measured for KMT-2022-BLG-2480 
and KMT-2024-BLG-0412.

\begin{figure}[t]
\includegraphics[width=\columnwidth]{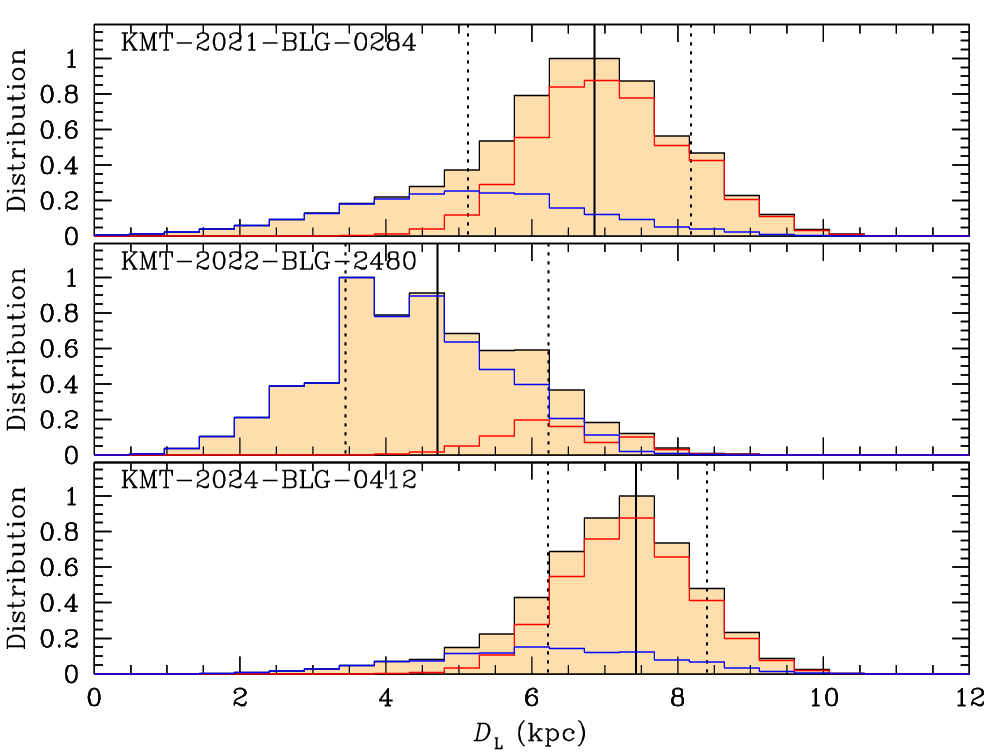}
\caption{
Bayesian posteriors of the distance to the lens. Notations are same as those in Fig.~\ref{fig:eight}.
}
\label{fig:nine}
\end{figure}

\begin{table}[t]
\caption{Physical lens parameters. \label{table:seven}}
\begin{tabular*}{\columnwidth}{@{\extracolsep{\fill}}lllcc}
\hline\hline
\multicolumn{1}{c}{Parameter}           &
\multicolumn{1}{c}{KB-21-0284}   &
\multicolumn{1}{c}{KB-22-2480}   &
\multicolumn{1}{c}{KB-24-0412}   \\
\hline
 $M_1$ ($M_\odot$)   &  $0.27^{+0.29}_{-0.17} $   &  $0.83^{+0.55}_{-0.34}$  &  $0.27^{+0.35}_{-0.14}$    \\  [0.6ex]
 $M_2$ ($M_\odot$)   &  $0.10^{+0.11}_{-0.06} $   &  $0.29^{+0.19}_{-0.12}$  &  $0.17^{+0.22}_{-0.09}$    \\  [0.6ex]
 $\dl$ (kpc)         &  $6.86^{+1.33}_{-1.73} $   &  $4.71^{+1.52}_{-1.27}$  &  $7.43^{+0.97}_{-1.21}$    \\  [0.6ex]
 $a_{\perp}$ (AU)    &  $1.81^{+0.35}_{-0.46} $   &  $5.11^{+1.65}_{-1.37}$  &  $5.94^{+0.78}_{-0.96}$    \\  [0.6ex]
 $p_{\rm disk}$      &   32\%                     &   88\%                   &   23\%                     \\  [0.6ex]
 $p_{\rm bulge}$     &   68\%                     &   12\%                   &   77\%                     \\
\hline
\end{tabular*}
\end{table}

We determine the physical lens parameters through a Bayesian analysis using priors on the physical
and dynamical distributions, as well as the mass function of lens objects within the Galaxy.  Using
these priors, we generated numerous artificial lensing events, each assigned a set of physical 
parameters $(M, \dl, \ds, \mu)_i$ from a Monte Carlo simulation. For the simulation, we employed 
the Galaxy model detailed in \citet{Jung2021} and the mass function model described in \citet{Jung2018}. 
For each artificial event, we compute the lensing observables  $(\te, \thetae)_i$ using the relations 
in Eq.~(\ref{eq5}).  We then constructed the posteriors for the lens mass and distance by assigning 
a weight to each artificial event.  The weight was calculated as 
\begin{equation}
w_i=\exp\left(-{\ \ \chi^2\over 2}\right); \qquad
\chi_i^2 = 
{(t_{{\rm E},i}-\te)^2\over \sigma(\te)^2} + 
{(\theta_{{\rm E},i}-\thetae)^2\over \sigma(\thetae)^2}.
\label{eq6}
\end{equation}
Here, $(\te, \thetae)$ denote the measured values of the observables, while $[\sigma(\te), 
\sigma(\thetae)]$ represent their respective measurement uncertainties.

Figures~\ref{fig:eight} and \ref{fig:nine} display the posteriors for the primary lens mass and the 
distance to the lens systems. The estimated values for the primary and companion lens masses, the 
distance to the lens, and the projected separation ($a_\perp$) between $M_1$ and $M_2$ for each 
event are listed in Table~\ref{table:seven}. Based on these parameters, it was found that the lenses 
of KMT-2021-BLG-0284 and KMT-2024-BLG-0412 are binaries composed of M dwarfs. For KMT-2022-BLG-2480, 
the primary lens is an early K-type main-sequence star, and the companion is an M dwarf. The table 
also includes the probabilities of the lens being in the disk ($p_{\rm disk}$) or the bulge 
($p_{\rm bulge}$). It was found that the lenses of KMT-2021-BLG-0284 and KMT-2024-BLG-0412 are 
likely located in the bulge, while the lens of KMT-2022-BLG-2480 is more likely in the disk.

\section{Summary and conclusion} \label{sec:six}

We conducted a project involving the systematic analysis of microlensing data from the KMTNet 
survey.  The objective of the project is to identify lensing events characterized by intricate 
anomaly features that are challenging to explain using conventional binary-lens or binary-source 
models.

Our investigation reveals that the light curves of microlensing events KMT-2021-BLG-0284, 
KMT-2022-BLG-2480, and KMT-2024-BLG-0412 exhibit highly complex patterns with three or more 
anomaly features. While these features could not be fully explained by a binary-lens (2L1S) 
model alone, the 2L1S model could effectively describe certain segments of the 
light curve. By incorporating an additional source into the modeling, we identified a 
comprehensive model that accounts for all the observed anomaly features.

Bayesian analysis, based on constraints provided by the lensing observables of the event 
timescale and the angular Einstein radius, indicates that the lenses of KMT-2021-BLG-0284 and 
KMT-2024-BLG-0412 are binary systems composed of two M dwarf stars. For KMT-2022-BLG-2480, the 
primary lens is an early K-type main-sequence star and the companion is an M dwarf. The binary 
stars in the sources for all events exhibit similar magnitudes due to a detection bias that 
favors binary source events featuring a relatively bright secondary star.

\begin{acknowledgements}
This research has made use of the KMTNet system operated by the Korea Astronomy and Space Science 
Institute (KASI) at three host sites of CTIO in Chile, SAAO in South Africa, and SSO in Australia. 
Data transfer from the host site to KASI was supported by the Korea Research Environment Open NETwork 
(KREONET).  This research was supported by the KASI under the 
R\&D program (Project No. 2023-1-832-03) supervised by the Ministry of Science and ICT.
The MOA project is supported by JSPS KAKENHI Grant Number JP24253004, JP26247023, JP23340064, 
JP15H00781, JP16H06287, JP17H02871 and JP22H00153.
J.C.Y., I.G.S., and S.J.C. acknowledge support from NSF Grant No. AST-2108414. 
Y.S.  acknowledges support from NSF Grant No. 2020740.
C.R. was supported by the Research fellowship of the Alexander von Humboldt Foundation.
W.Z. and H.Y. acknowledge support by the National Natural Science Foundation of China (Grant
No. 12133005).  W. Zang acknowledges the support from the Harvard-Smithsonian Center for Astrophysics 
through the CfA Fellowship.
\end{acknowledgements}

\end{document}